\documentclass[aps,pra,twocolumn,showpacs,groupedaddress]{revtex4}
\usepackage{mathrsfs}
\usepackage{amsfonts}
\usepackage{amssymb}
\usepackage{graphicx}

\DeclareGraphicsExtensions{.eps,.mps,.pdf,.jpg,.png}
\begin{document}
\title{Anti-adiabatic evolution in quantum-classical hybrid system}

\author{J. Shen}
\affiliation{Fundamental Education Department,  Dalian Neusoft University of Information, Dalian 116023,  China}

\author{W. Wang}
\affiliation{%
School of Physics, Northeast Normal University, Changchun 130024, China}

\author{C. M. Dai}%
\affiliation{%
Center for Quantum Sciences and School of Physics, Northeast Normal University, Changchun 130024, China}%

\author{X. X. Yi}
\altaffiliation{yixx@nenu.edu.cn}
\affiliation{%
Center for Quantum Sciences and School of Physics, Northeast Normal University, Changchun 130024, China}%

\date{\today}

\begin{abstract}
The adiabatic theorem is an important concept in quantum mechanics,
it tells that a quantum  system subjected to gradually changing
external conditions remains to the same instantaneous eigenstate of
its Hamiltonian as it initially in. In this paper, we study the
another extreme circumstance where the external conditions  vary
rapidly such that  the quantum system can not follow the  change and
remains in its initial state (or wavefunction). We call this type of
evolution anit-adiabatic evolution.  We examine the matter-wave
pressure in this situation and derive the condition for such an
evolution. The study is conducted by considering a quantum particle
in an infinitely deep potential, the potential width $Q$ is assumed
to be change rapidly. We show that the total energy of the quantum
subsystem  decreases as $Q$ increases, and this rapidly change
exerts a force on the wall which plays the role  of boundary of the
potential. For $Q<Q_{0}$ ($Q_0$ is the initial width of the
potential), the force is repulsive, and for $Q>Q_{0}$, the force is
positive. The condition for the anti-adiabatic evolution is given via a spin-$\frac 1 2$ in a rotating magnetic field.
\end{abstract}

 \maketitle

\section{introduction}
A quantum  system would remain in the instantaneous eigenstate of
its Hamiltonian if the Hamiltonian changes slowly enough with
respect to the energy gaps among the instantaneous eigenstate
\cite{P. Ehrenfest16,M. Born28,J. Schwinger37, T. Kato50}. This is
the so-called adiabatic theorem, which is an important and intuitive
concept in quantum mechanics, and it was found insightful and
potential to applications. For example, Landau-Zener transition,
Berry phase \cite{L. D. Landau32, C. Zener32, M. Gell-mann51, M. V.
Berry84}, quantum control and quantum adiabatic
computation\cite{T.Corbitt07, M.Bhattacharya08}. Although progresses
have been made,   there are many issues remain open concerning the
adiabatic evolution, for instance, the adiabatic condition \cite{K.
P. Marzlin04, D. M. Tong05} and its extension to open systems
\cite{yijpb07}.

Recent years have witnessed a series of developments at the
intersection of optical cavities and  mechanical
resonators\cite{craighead00,aspelmeyer14}. The opto-mechanical coupling between a
moving mirror and the radiation pressure of light has first appeared
in the context of interferometric gravitational wave experiments.
Owing to the discrete nature of photons, the quantum fluctuations of
the radiation pressure forces give rise to the so-called standard
quantum limit\cite{caves81}.   The experimental manifestations of
opto-mechanical coupling by radiation pressure have been observable
for some time. For instance, radiation pressure forces were observed
in\cite{dorsel83}, while even earlier work in the microwave domain
had been carried out by Braginsky\cite{braginsky77}. Moreover the
modification of mechanical oscillator stiffness caused by radiation
pressure, the \textbf{optical spring}, have also recently been
observed\cite{sheard04}.

It is the similarity between light and matter-wave that motivates
the concept so-called matter-wave pressure\cite{J.Shen10}. By
examining  the dynamics of the adiabatic quantum-classical system,
the authors calculated the force exerted on the classical subsystem
by the quantum subsystem\cite{J.Shen10}. In the analysis,   an
assumption that the classical system moves slowly is used, this
leads to the adiabatic evolution for  the quantum subsystem.

On the contrary, quantum quenching refers to a sudden change of some parameter of the Hamiltonian. A variety of processes can result in quenching, such as a sudden  moving of the mirror in optomechanics, a spin rotating driven by a magnetic field. Recently, the concept of phase transitions has been extended to non-equilibrium dynamics of time-independent systems induced by a quantum
quench\cite{arkadiusz17}. It has been shown that the quantum quench in a discrete time crystal leads to dynamical quantum phase transitions, and  the return probability of a periodically driven system to a Floquet eigenstate before the quench reveals singularities in time.
Based on random quenches,  random unitaries in atomic Hubbard and spin models can be generated\cite{elben17}, this proposal works for a broad class of atomic and spin lattice models\cite{vermersch18}.

It is believe that  in the case of rapidly varying conditions (to
which the quantum system subjected),  the quantum system  may has no
time to change its state\cite{citro05,pellegrini11,eidelstein13}. If this is the case, what is the condition
for such an evolution? Is it that the quantum system has no time to
change its state? or it can not follow the rapidly changing
conditions?  How the matter-wave pressure behaves in this
circumstance? In this paper, we will focus on these questions.

The paper is organized as follows. In Sec.{\rm II}, we study the
dynamics of a quantum particle in an infinitely deep potential with
varying potential width.  The matter-wave pressure force is
calculated and discussed by assuming that the system remains in its
initial state. In Sec.{\rm III}, we derive the condition of
anti-adiabatic evolution trough a simple example. Finally, we
conclude our results in Sec.{\rm IV}.

\section{a quantum particle  in an infinite one-dimensional potential with varying width}
\begin{figure}[t]
\centering
\includegraphics*[width=0.7\columnwidth,
height=0.5\columnwidth]{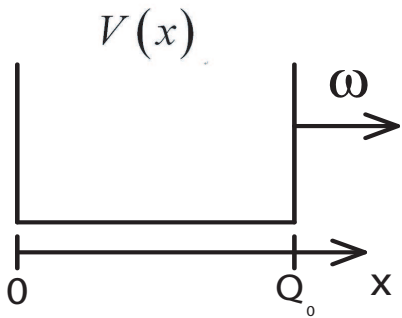} \caption{A infinitely deep potential
well with a moving wall as its \textbf{right} boundary, the wall may
move fast with respect to other parameters in the system, in
particular the time scale of transitions among the
particle-potential energy levels.} \label{fig1}
\end{figure}

Consider  a quantum system in a one-dimensional  infinitely deep
well whose boundary at $Q$  is a moving wall(see Fig.\ref{fig1}).
The Hamiltonian  $\hat{H}$ of this system can be written as,
\begin{eqnarray}
\hat{H}=-\frac{\hbar^{2}}{2m}\frac{d^{2}}{dx^{2}}+V(x),\label{Hamiltonian}
\end{eqnarray}
where the potential $V(x)$ takes,
\begin{eqnarray}
V(x)=\left\{\begin{array}{cc}
       0 & 0\leq x \leq Q \\
       +\infty &  x<0 ~\text{or} ~x>Q
     \end{array}\right. .
\end{eqnarray}
Suppose that  the moving wall at $Q$ only changes the boundary
condition of the quantum system, we have the eigenvalues and the
corresponding  eigenstates of the quantum system with fixed $Q$,
\begin{eqnarray}
&&\Psi_n(Q)=\langle q|\psi_{n}(q, Q)\rangle=\sqrt{\frac{2}{Q}}\sin\frac{n\pi q}{Q}, \nonumber\\
&&E_{n}(Q)=\frac{\hbar^{2}\pi^{2}n^{2}}{2mQ^{2}},   ~~~~~n=1,2,3\cdots,  \label{eigenvalue}
\end{eqnarray}
where $q$ denotes the coordinate of the quantum particle.

In the following, we assume that the  wall moves so fast such that
the quantum system does not evolve. The condition for this
assumption to hold will be discussed in the next section. Suppose
that the quantum particle is initially prepared in the ground state
with the boundary at $Q_0$, i.e.,
\begin{eqnarray}
&&\Psi_{1}(Q_0)=\sqrt{\frac{2}{Q_{0}}}\sin\frac{\pi q}{Q_{0}}, \nonumber\\
&&E_1(Q_0)=\frac{\hbar^{2}\pi^{2}}{2mQ_{0}^{2}}.
\end{eqnarray}
At next instance of time, the wall moves to $Q$. Since we assume the
wall to moves so fast, the particle does not evolve and  keeps in
$\Psi_1(Q_0)$ that can be expended as a function of $\Psi_{n}(Q)$,
\begin{eqnarray}
\Psi_{1}(Q_0)=\sum_{n} b_{n}\Psi_{n}(Q),
~~~~~n=1,2,3,\cdots,\label{assumption}
\end{eqnarray}
where $b_{n}$ is the expansion coefficients,   and we define
$\rho_{n}=|b_{n}|^{2}$, standing for  the probability of the
particle  in the $n$th energy-level with the wall at $Q$. Simple
calculation shows that,
\begin{eqnarray}
b_{n} =\left\{
        \begin{array}{ll}
          \frac{(-1)^{n}2n\sqrt{\gamma}\sin(\gamma\pi)}{\pi(\gamma^2-n^2)}, &~~ \gamma < 1; \\
          \frac{2\gamma^\frac{3}{2}}{\pi(\gamma^2-n^2)}
          \sin\frac{n\pi}{\gamma}, & ~~\gamma > 1 \text{ and } n \neq \gamma; \\
          \frac{1}{\sqrt{\gamma}}, & ~~n = \gamma.
        \end{array}
      \right.
\label{population}
\end{eqnarray}
where $\gamma=\frac{Q}{Q_{0}}$ was defined.  $\rho_{n}$ follows
from Eq.(\ref{population}),
\begin{eqnarray}
\rho_{n}=\left\{
           \begin{array}{ll}
\frac{4n^2\gamma\sin^2(\gamma\pi)}{\pi^2(\gamma^2-n^2)^2}, &~~ \gamma < 1; \\
\frac{4\gamma^3}{\pi^2(\gamma^2-n^2)^2}
\sin^2(\frac{n\pi}{\gamma}), & ~~\gamma > 1 \text{ and } n \neq \gamma; \\
             \frac{1}{\gamma}, & ~~n = \gamma.
           \end{array}
         \right.
\label{probability}
\end{eqnarray}

\begin{figure}[h]
\centering
\includegraphics*[width=1.1\columnwidth,
height=0.8\columnwidth]{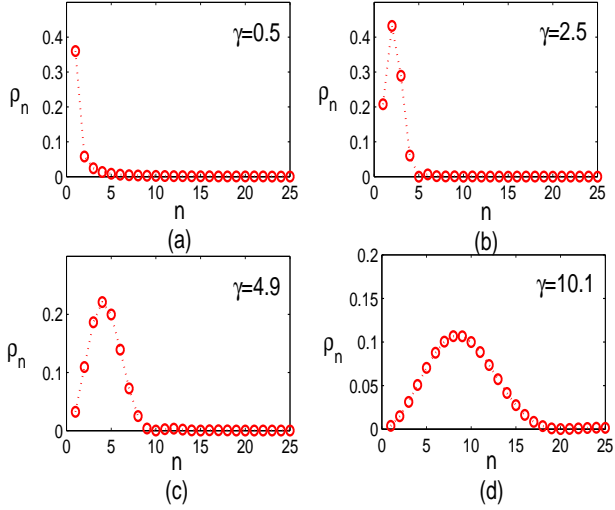} \caption{(Color on line)~The
population $\rho_{n}$ as a function of n, where the energy level is
calculated with the boundary at $Q$. Where $\gamma$ are taken as
0.5, 1.5, 4.9 and 10.1 respectively.} \label{fig2}
\end{figure}

\begin{figure}[h]
\centering
\includegraphics*[width=0.8\columnwidth,
height=0.9\columnwidth]{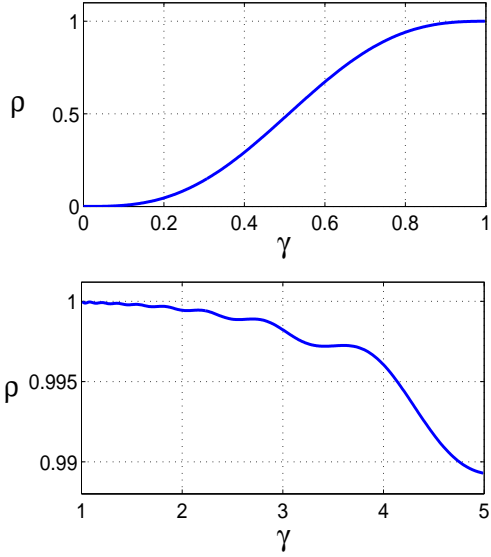}
\caption{(Color on line)~The total probability of the first ten energy-levels.}
\label{fig3}
\end{figure}

From Eq.(\ref{probability}), we find that the probability $\rho_{n}$
is only a function of $\gamma$ and $n$, indicating that  $Q$ and
$Q_0$ jointly changes $\rho_{n}$. Furthermore, expression for
$0<\gamma<1$ and $\gamma>1$ is different. In Fig.\ref{fig2} we plot
the probability distribution over the energy level $\Psi_n(Q)$,
while $\gamma$ is chosen to 0.5, 1.5, 4.9 and 10.1,  respectively.
From this figure, we  find that there are population transfer from
the ground state to the higher energy levels that is different from
the results in Ref.\cite{J.Shen10}. We also  find that the
probability distribution sharply depends on $\gamma$. For example,
when $\gamma=4.9$,  the particle  mainly occupied the 5th level,
while the occupation  of the other levels, especially that far from
the 5th levels, are almost zero. From Eq.(\ref{eigenvalue}), we
observe that the $n$th eigenfunction with boundary at $Q$ is similar
to the initial state when $n \approx \gamma$. This may be the reason
why the probability of the energy-level with index (i.e., $n$) close
to $\gamma$ is favoringly  occupied.

To calculate the energy change   due to boundary moving, we have to
calculate the population distribution over all eigenstates of the
Hamiltonian with the new boundary. This is a time-consuming task.
Fortunately,  our calculation show that only first 10 levels are
populated when the boundary change is from $0$ to
$5Q_{0}$(see Fig.\ref{fig3}). Then we can only take the first ten
energy-levels into account, which is a good approximation to
calculating the energy for the parameter under our discussion.

Since $\Psi_{1}$ can be expended as Eq.(\ref{assumption}),  we can
easily get the total energy after the boundary moving to the new
position,
\begin{eqnarray}
E^{'}&=&\sum_{n=1}^{10} \rho_{n}^{'}E_{n}^{'}\nonumber\\
&=&\sum_{n=1}^{10} \frac{\rho_{n}}{\sum_{n=1}^{10}\rho_{n}}E_{n}^{'}\nonumber\\
&=&\sum_{n=1}^{10} \frac{\rho_{n}}{\sum_{n=1}^{10}\rho_{n}}
\frac{\hbar^2\pi^2n^2}{2m(\gamma Q_{0})^2}\nonumber\\
&=&\sum_{n=1}^{10} \frac{\rho_{n}}{\sum_{n=1}^{10}\rho_{n}}
\frac{n^2}{\gamma^2}\cdot E_{1}, \label{energy'}
\end{eqnarray}
where $\rho^{'}$ is the re-normalized probability distribution over
the first ten energy-levels.

\begin{figure}
\centering
\includegraphics*[width=0.9\columnwidth,
height=0.6\columnwidth]{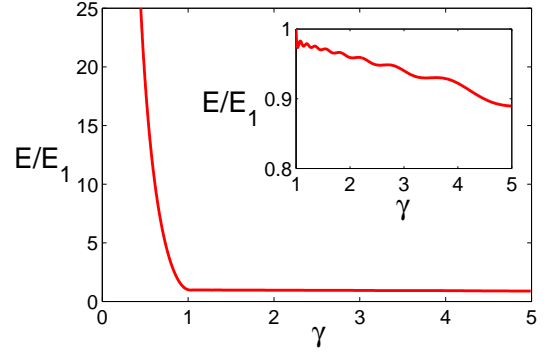} \caption{(Color on line)~The energy
(in units of $E_{1}$) of the atom as a function of the ratio
$\frac{Q}{Q_{0}}$ with the initial condition $Q_{0}=1.0$ (in units
of nm) and
$E_{1}=5.49*10^{-23}j$($m=10^{-27}kg,h=6.626*10^{-34}j\cdot s$).}
\label{fig4}
\end{figure}

\begin{figure}
\centering
\includegraphics*[width=0.9\columnwidth,
height=0.6\columnwidth]{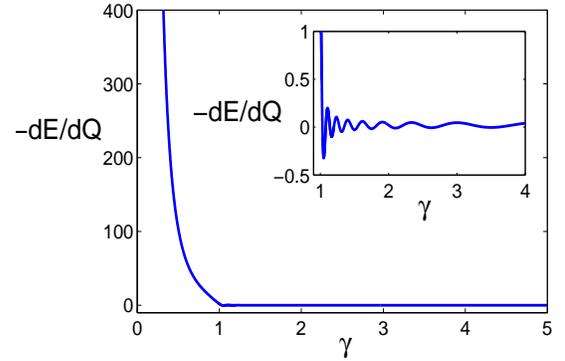} \caption{(Color on line)~The force(in
units of $\frac{E_{1}}{Q_{0}}$) between the quantum subsystem and
the classical subsystem with $Q$ is in units of nm and the energy
$E$ is in units of $E_{1}$, where
$E_{1}=5.49*10^{-23}j$($m=10^{-27}kg,h=6.626*10^{-34}j\cdot s$).}
\label{fig5}
\end{figure}

\begin{figure}
\centering
\includegraphics*[width=0.8\columnwidth,
height=0.9\columnwidth]{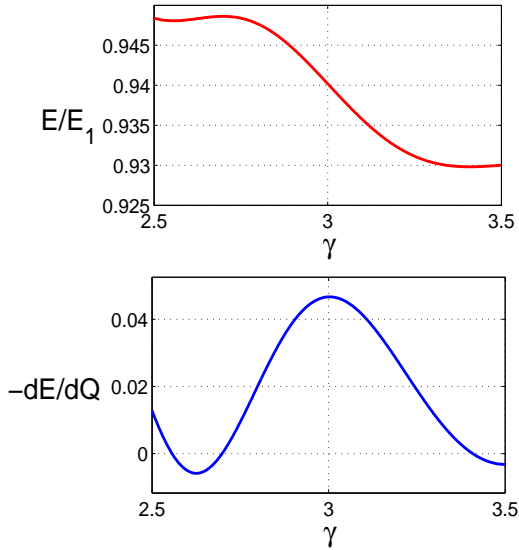} \caption{(Color on line)~A
local graphics amplification of Fig.\ref{fig4} and Fig.\ref{fig5}
from $\gamma=2.5$ to $\gamma=3.5$.} \label{fig6}
\end{figure}

From Eq.(\ref{probability}) and Eq.(\ref{energy'}),  we find that
$E^{'}$ depends only on $\gamma$ and $n$. In other words, $E^{'}$
does not depends on $Q$ and $Q_0$ separately. So it is interesting
to see how the ratio $\gamma$ affects the total energy when the
boundary changes rapidly.

Fig.\ref{fig4} shows the dependence of the energy on $\gamma$.  We
find that the total energy increases rapidly as $\gamma$ decreases
in the regime $0<\gamma<1$. This observation can be easily
understood by examining Eq.(\ref{eigenvalue}). It is obviously that
$E_{n}$ is inversely proportional to the square of the width of the
well $Q$, which means that the energy of each energy-level increases
as  $Q$ decreases and  $Q<Q_{0}$. Moreover, because we choose the
ground state as the initial state,  no matter  in what state the
particle will be after the boundary change, the total energy will
certainly increases.

For  $\gamma>1$, the situation is different. We find that the total
energy decreases at large  as $\gamma$ increases. The total energy
change is almost zero when the change of $Q$ is not very large.
Meanwhile we find that the energy does change monotonically with
$\gamma$. In other words, the energy may increase as $\lambda$
increases. This is interesting. In Ref.\cite{J.Shen10}, the total
energy decreases as $\gamma$ increases. This is because the
evolution of the  system is adiabatic and the particle  is always at
the ground state of the Hamiltonian, no matter how the boundary
changes. In the other words, the width of the well $Q$ is the only
parameter to determine the energy of the quantum subsystem. However,
this is not the case for anti-adiabatic evolution  in our model.
Indeed, there are population transfer among the eigenstates when the
boundary changes from $Q_{0}$ to $Q$, see Eq.(\ref{assumption}).
Namely the particle will not always stay in the ground state when
the boundary changes. This will  affect the total energy of the
system. From Eq.(\ref{energy'}), we can see that the total energy is
related to $b_{n}$, which varies  as the boundary changes. This
analysis suggests that the energy depends on two parameters, one the
eigenvalue $E_{n}$ and the other is the population distribution
$\rho_{n}$. From Eq.(\ref{eigenvalue}), we can see that when  $Q$
increases,  the eigenvalue $E_{n}(Q)$ decreases. On the other hand,
from Eqs.(\ref{population}) and (\ref{probability}), we see that the
change of $\gamma$ will result in the change of the population
$\rho_{n}$. Specifically, we find that the probability distribution
mainly in a few energy levels near $\gamma$ for $\gamma>1$. These
together can interpret why  the total energy increase as $Q$
increases. see Fig.{\ref{fig4}.

Now we discuss this issue from the aspect of matter wave  force
exerted on the boundary wall.  It can be given by $F=-\frac{dE}{dQ}$
\cite{J.Shen10}. We show this force in Fig. \ref{fig5}. From
Fig.\ref{fig5}, we find that the force tends to very large and
repulsive  as $Q$ decreases in the regime $Q<Q_{0}$. This is similar
to the result in Ref.\cite{J.Shen10}. In addition, for the case of
$Q>Q_{0}$,  the force has a slight fluctuation around zero with the
increasing of $Q$. This means that the force between the particle
and the moving wall may be repulsive or  attractive. Fig.\ref{fig6}
is a enlarged version of Fig.\ref{fig4} and Fig.\ref{fig5} for
$\gamma$ ranging from $\gamma=2.5$ to $\gamma=3.5$. From the two
figures, we observe that when the total energy increases, the force
is  attractive. On the contrary, there is a repulsive force when the
total energy decreases. Since the change  of the total energy due to
the boundary moving is so weak for $Q>Q_{0}$, the force in this case
is negligibly small.

\section{a spin-$\frac 1 2$ in a rotating magnetic  field}
In the last section, we study the matter-wave pressure with an
assumption that the boundary moves so fast that the quantum system
does not evolve. One may wonder if this situation exist, and what is
the condition for such an evolution. Does the system have no time to
evolve? Or the change is too fast that the system can not follow? To
simplify the discussion, we here adapt a simple model that a
spin-$\frac 1 2$ in a rotating magnetic field to formulate the
problem.

The system Hamiltonian takes, $\hat{H}=-\vec{\mu} \cdot \vec{B(t)}$.
We will choose $\vec{B(t)}=B_{0}\widehat{n(t)}$ with the unit vector
$\widehat{n(t)}=(\sin \alpha \cos \omega t,\sin \alpha \sin \omega
t,\cos \alpha)$ as the magnetic field, where $B_{0}$ is strength of
the field. The eigenvalue and the corresponding  eigenstate of the
system takes,
\begin{eqnarray}
\mathbf{\psi_{1}}(t)&=&(\cos(\alpha/2), e^{i\omega t}\sin(\alpha/2))^{T} , \nonumber\\
E_{1}&=&+\frac{\hbar\omega_{0}}{2} \label{stateequation1}
\end{eqnarray}
and
\begin{eqnarray}
\mathbf{\psi_{2}}(t)&=&(e^{-i\omega t}\sin(\alpha/2),-\cos(\alpha/2))^{T} , \nonumber\\
E_{2}&=&-\frac{\hbar\omega_{0}}{2}
\end{eqnarray}
where $E_{1}$ and $E_{2}$ are the eigenvalues of $\mathbf{\psi_{1}}$
and $\mathbf{\psi_{2}}$, respectively. And $\omega$ is the frequency
of the magnetic field, $\alpha$ denotes the angle between the spin
and the magnetic field. $\omega_{0}\equiv\frac{eB_{0}}{m}$, and  $e$
is the charge of the particle,  $m$ is the mass of the particle.
Starting with $\mathbf{\psi_{1}}(t=0)$, the particle will evolve to
\begin{equation}
\mathbf{\psi}(t)=\left(
                   \begin{array}{c}
                     ( \cos(\frac{\lambda t}{2})-i(\frac{\omega_{0}-\omega}{\lambda})
                     \sin(\frac{\lambda t}{2}) )\cos(\frac{\alpha}{2})
                     e^{-\frac{i\omega t}{2}} \\
                     (\cos(\frac{\lambda t}{2})-i(\frac{\omega_{0}+\omega}{\lambda})
                     \sin(\frac{\lambda t}{2}))\sin(\frac{\alpha}{2})
                     e^{\frac{i\omega t}{2}} \\
                   \end{array}
                 \right)
\end{equation}
where $\lambda$ is defined by,
\begin{equation}
\lambda=\sqrt{\omega^{2}+\omega_{0}^{2}-2\omega\omega_{0}\cos(\alpha)}.
\end{equation}
We now examine in which circumstance the system remains un-evolved
on $\mathbf{\psi_{1}(0)}$. This can be done by calculating the
probability of the particle on $\mathbf{\psi_{1}(0)}$,
\begin{eqnarray}
\rho_{1}(t)&=&|\langle\mathbf{\psi}(t)|\mathbf{\psi_{1}(0)}\rangle|^{2}  \nonumber\\
&=&[\cos(\frac{\lambda t}{2}) \sin(\frac{\omega t}{2}) \cos\alpha  \nonumber\\
&&+ (\frac{\omega_0 - \omega \cos\alpha}{\lambda})\sin(\frac{\lambda t}{2}) \cos(\frac{\omega t}{2})]^{2}  \nonumber\\
&&+ [\cos(\frac{\lambda t}{2}) \cos(\frac{\omega t}{2}) \nonumber\\
&&+ (\frac{\omega - \omega_0 \cos\alpha}{\lambda})\sin(\frac{\lambda t}{2}) \sin(\frac{\omega t}{2})]^{2}.
\label{pro1}
\end{eqnarray}
We are interested in the probability of the particle in its initial
state, when the magnetic filed completes a circle. Eq.(\ref{pro1})
would give the results if we assume that the evolution time $t$ and
the magnetic frequency $\omega$ satisfy $\omega t=2\pi$. With
$\omega t=2\pi$, Eq.(\ref{pro1})  can be rewritten as
\begin{eqnarray}
\rho_1(\omega)&=&\cos^2(\frac{\lambda\pi}{\omega})
+(\frac{\omega_0-\omega \cos\alpha}{\lambda})^2\sin^2(\frac{\lambda\pi}{\omega}).
\label{rho1}
\end{eqnarray}

On the contrary, the particle will evolve to $\mathbf{\psi^{'}(t)}$ given blow when starting with $\mathbf{\psi_{2}}(t=0)$,
\begin{equation}
\mathbf{\psi}^{'}(t)=\left(
                   \begin{array}{c}
                     (\cos(\frac{\lambda t}{2})+i(\frac{\omega_{0}+\omega}{\lambda})
                     \sin(\frac{\lambda t}{2}))\sin(\frac{\alpha}{2})
                     e^{-\frac{i\omega t}{2}} \\
                     -(\cos(\frac{\lambda t}{2})+i(\frac{\omega_{0}-\omega}{\lambda})
                     \sin(\frac{\lambda t}{2}))\cos(\frac{\alpha}{2})
                     e^{\frac{i\omega t}{2}} \\
                   \end{array}
                 \right)
\end{equation}

\begin{eqnarray}
\rho_{2}^{'}(t)&=&|\langle\mathbf{\psi}^{'}(t)|\mathbf{\psi_{2}(0)}\rangle|^{2}  \nonumber\\
&=&[\cos(\frac{\lambda t}{2}) \sin(\frac{\omega t}{2}) \cos\alpha  \nonumber\\
&&+ (\frac{\omega_0 - \omega \cos\alpha}{\lambda})\sin(\frac{\lambda t}{2}) \cos(\frac{\omega t}{2})]^{2}  \nonumber\\
&&+ [\cos(\frac{\lambda t}{2}) \cos(\frac{\omega t}{2}) \nonumber\\
&&+ (\frac{\omega - \omega_0 \cos\alpha}{\lambda})\sin(\frac{\lambda t}{2}) \sin(\frac{\omega t}{2})]^{2}.
\label{pro2}
\end{eqnarray}

\begin{eqnarray}
\rho_{2}^{'}(\omega)&=&\cos^2(\frac{\lambda\pi}{\omega})
+(\frac{\omega_0-\omega \cos\alpha}{\lambda})^2\sin^2(\frac{\lambda\pi}{\omega}).
\label{rho2}
\end{eqnarray}
From Eqs(\ref{pro2}) and (\ref{rho2}), we  observe that the expressions of $\rho_{2}^{'}(t)$ and $\rho_{2}^{'}(\omega)$ both are the same as $\rho_{1}(t)$ and $\rho_{1}(\omega)$. Hence, we only discuss the evolution of $\psi(t)$ in the following discussions.

\begin{figure}[h]
\centering
\includegraphics*[width=0.8\columnwidth,
height=0.6\columnwidth]{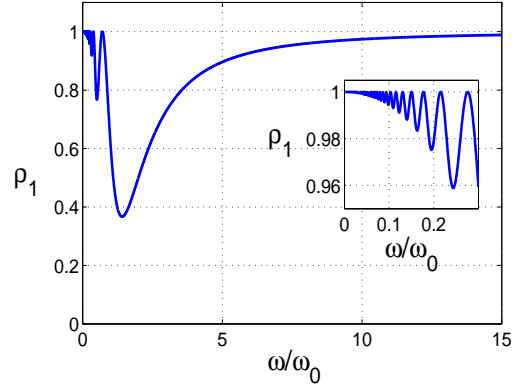} \caption{(Color on line)~The
relationship between $\rho_1$ and $\omega$ with
\textbf{$\alpha=\frac{\pi}{4}$,}
$\omega_{0}\equiv\frac{\mathbf{e}B_{0}}{m}$ being a c-frequency and
$B_{0}=1 ~T$, where $e ~(-1.6\times10^{-19} ~c)$ is the charge of
the electron and $m ~(9.3\times10^{-31} ~kg)$ is the  mass of the
electron.} \label{omega1}
\end{figure}

In Fig.\ref{omega1}, we plot $\rho_{1}$ as a function of $\omega$
for $\alpha=\frac{\pi}{4}$. From this figure, we can  find that when
$\frac{\omega}{\omega_{0}}$ is very small, the evolution of
$\rho_{1}$ is irregular, for this reason we can not find a suitable
frequency $\omega$ to make sure that the electron will stay in the
initial state. Fortunately, $\rho_1$ increases with the increasing
of $\omega$ when $\frac{\omega}{\omega_{0}}>1.442$ and we find that
when $\omega$ is 15 times larger than $\omega_0$, the probability
$\rho_1$ is almost one and we claim  that the electron will stay at
the state $\mathbf{\psi_1}(0)$ at any time. This suggests  that the
quantum system will stay in its initial state if the external
conditions change much faster than the typical frequency of the
system.

\begin{figure}[h]
\centering
\includegraphics*[width=0.8\columnwidth,
height=0.6\columnwidth]{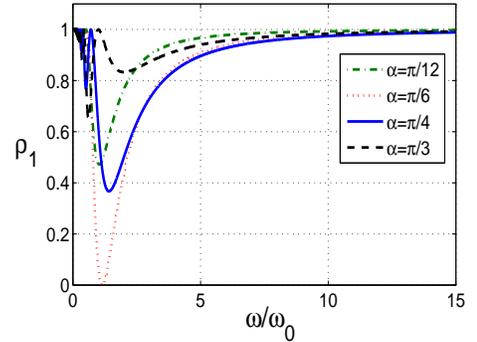} \caption{(Color on line)~The
relationship between $\rho_1$ and $\omega$ with
$\alpha=\frac{\pi}{12},\frac{\pi}{6},\frac{\pi}{4}$ and
$\frac{\pi}{3}$, respectively. And
$\omega_{0}\equiv\frac{\mathbf{e}B_{0}}{m}$ being a c-frequency and
$B_{0}=1 ~T$, where $e ~(-1.6\times10^{-19} ~c)$ is the charge of
the electron and $m ~(9.3\times10^{-31} ~kg)$ is the  mass of the
electron.}\label{omegaa}
\end{figure}

In Fig.\ref{omegaa}, we plot $\rho_{1}$ as a function of
$\omega$ for different $\alpha$. From this figure, we find that the
minimum value of $\rho_{1}$ changes for different $\alpha$. However,
there always exists a $\omega$ which can keep the system at the state $\mathbf{\psi_1}(0)$ at any time, no matter
what  $\alpha$ is. And  $\omega$ for different $\alpha$ is
nearly the same.

\section{Conclusion and discussions}

The adiabatic theorem tells that a quantum mechanical system
subjected to gradually changing external conditions can adapt its
functional form. In this paper, we explore another extreme varying
conditions--rapidly varying conditions. The evolution of the system
in this condition we call anti-adiabatic evolution. We have examined
the condition for such evolutions  and calculate the matter-wave
pressure for the quantum system. Specifically, we have considered a
quantum particle in a one-dimensional infinitely deep potential, one
boundary of the potential is assumed to move rapidly, such that the
particle inside does not evolve with time, however, as the potential
width varies, the energy of the particle changes. This change would
lead to a force on the quantum system. We calculated the force and
find that as the width increases the force is attractive, while it
is repulsive as the width decreases. By considering a spin-$\frac 1
2 $ in a rotating magnetic field, we explore the condition for the
anti-adiabatic evolution. Discussions and remarks on this condition
are given.

\end{document}